\documentclass[twocolumn,amsmath,amssymb,prc,superscriptaddress,floatfix,showpacs]{revtex4}

\usepackage{graphicx} \usepackage{color} \usepackage{hyperref}

\newcommand{\bj}{\marginpar{\colorbox{green}{\textbf{BJ}}\\@MW}}

\def\roughly#1{\mathrel{\raise.3ex\hbox{$#1$\kern-.75em%
\lower1ex\hbox{$\sim$}}}}

\newcommand{\hlowk}{\ensuremath{\mathcal{H}_{\text{low }k}}}
\newcommand{\vlowk}{\ensuremath{V_{\text{low }k}}}

\newcommand{\ket}[1]{\left|{#1}\right\rangle}
\newcommand{\bra}[1]{\left\langle{#1}\right|}
\newcommand{\norm}[1]{\ensuremath{\left\Vert #1 \right \Vert}}

\newcommand{\einh}[1]{\ensuremath{\,\text{#1}}}
\newcommand{\MeV}{\einh{MeV}} 
\newcommand{\diag}{\ensuremath{\operatorname{diag}}}

\graphicspath{{./}}

\begin{document}
\title{On the convergence of multi-channel effective interactions} \author{M. Wagner}
\email[E-Mail:]{mathias.wagner@physik.tu-darmstadt.de}
\affiliation{Institut f\"{u}r Kernphysik, TU Darmstadt, D-64289
  Darmstadt, Germany} \author{B.-J. Schaefer}
\email[E-Mail:]{bernd-jochen.schaefer@uni-graz.at}
\affiliation{Institut f\"{u}r Physik, Karl-Franzens-Universit\"{a}t,
  A-8010 Graz, Austria} \affiliation{Institut f\"{u}r Kernphysik, TU
  Darmstadt, D-64289 Darmstadt, Germany} \author{J. Wambach}
\affiliation{Institut f\"{u}r Kernphysik, TU Darmstadt, D-64289
  Darmstadt, Germany} \affiliation{Gesellschaft f\"{u}r
  Schwerionenforschung GSI, D-64291 Darmstadt, Germany} \author{T.T.S.
  Kuo} \author{G.E. Brown} \affiliation{Department of Physics and
  Astronomy, State University of New York, Stony Brook, NY 11794-3800,
  USA}

\date{\today}


\begin{abstract}
  A detailed analysis of convergence properties of the
  Andreozzi-Lee-Suzuki iteration method, which is used for the
  calculation of low-momentum effective potentials $\vlowk$ is
  presented. After summarizing different modifications of the
  iteration method for one-flavor channel we introduce a simple model
  in order to study the generalization of the iteration method to
  multi-flavor channels. The failure of a straightforward
  generalization is discussed. The introduction of a channel-dependent
  cutoff cures the conceptual and technical problems. This novel
  method has already been applied successfully for realistic
  hyperon-nucleon interactions.

\end{abstract}

\maketitle

\newpage

%
%
\section{Introduction}
\label{sec:intro}

In Ref.~\cite{SKB03} Bogner et al.~have developed a low-momentum $NN$
interaction $\vlowk$ which is constructed by a combination of
renormalization group (RG) ideas and effective field theory
approaches. Starting from a realistic nucleon-nucleon interaction in
momentum space for the two-nucleon system, the $\vlowk$ is obtained by
integrating out the high-momentum components of the potential. In this
way a cutoff momentum $\Lambda$ is introduced to specify the boundary
between the low- and high-momentum space. Since the physical
low-energy quantities such as phase shifts or the deuteron binding
energy must not depend on the cutoff, the full-on-shell $T$-matrix has
to be preserved for the relative momenta $k, k' < \Lambda$. This
yields an exact RG flow equation for a cutoff-dependent effective
potential $\vlowk$, which is based on a modified Lippmann-Schwinger
equation (LSE).

Instead of solving the RG flow equation, which is a differential
equation, with standard methods (e.g.~Runge-Kutta method) directly, an
iteration method is used. This iteration method, pioneered by
Andreozzi, Lee and Suzuki (ALS)~\cite{lee80,suzuki80,andreozzi}, is
based on a similarity transformation and its solution corresponds to
solving the flow equation~\cite{Bogner2001}.This transformation is also used for the effective interaction in the
no-core shell model (e.g.~\cite{Navratil:1999pw,Stetcu:2004wh}).

In Ref.~\cite{1letter} the $\vlowk$ approach was generalized to the
hyperon-nucleon ($YN$) interaction with strangeness $S=-1$. For
hyperons with strangeness $S=-1$, two isospin states $I=1/2$ and
$I=3/2$ are available. For $I=1/2$, several coupled hyperon-nucleon
channels occur that require new technical developments in the
iteration method as already outlined in Ref.~\cite{1letter}.
Especially, the $\Lambda N \leftrightarrow \Sigma N$ transition
involves a mass difference of $\approx 78 \MeV$ and causes the failure
of the known methods. In the present study we discuss these novel
technical modifications in detail and investigate the convergence of
the ALS iteration methods.

The paper is organized as follows: In the next Sec.\ we summarize the
ALS iteration method for the one-flavor channel and discuss several
modifications which improve its convergence. The generalization of the
ALS method to multi-flavor channels is presented in the following
Section. Instead of working with a realistic hyperon-nucleon
interaction we introduce a simple solvable model in order to elucidate
the essential modifications of the iteration method. We end with a
summary and outlook.

%
%
\section{One-flavor Channel}
\label{sec:oneflavor}

The so-called model space (or $P$-space) is introduced in order to
truncate the infinitely many degrees of freedom of a Hilbert space to
a finite, low-energy subspace. The projection operator $P$ onto the
model space and its complement projection operator $Q$ onto the
excluded space are defined by
\begin{equation}
  P \equiv \sum\limits_{i=1}^d \ket{\phi_i}\!\bra{\phi_i}
  \quad \ ; \qquad Q = 1-P\ ,
\end{equation}
where $d$ denotes the dimension of the model space. The wave functions
$\ket{\phi_i}$ are eigenfunctions of the unperturbed Hamiltonian $H_0$,
which is composed of the kinetic energy and an arbitrary one-body
potential.

Employing this separation of the Hilbert space, the original full
Hamiltonian $H$ can then be replaced by a transformed Hamiltonian
$\mathcal{H}$ that acts only on the low-energy model subspace. In
particular, if $\ket{\Psi_i}$ is an eigenstate of the full Hamiltonian
$H$ then its projection onto the model space
$\ket{\psi^M_i}=P\ket{\Psi_i}$ satisfies the Schr\"odinger equation
\begin{equation}\label{eq:psimodel}
  \mathcal{H}\ket{\psi^M_i}=
  E_i\ket{\psi^M_i}\ .
\end{equation}

The crucial point is that the eigenvalues $E_i$ of the transformed
Hamiltonian are the same as those of the original Hamiltonian.

On the other hand, the transformation of all model states
$\ket{\psi^M_i}$ back into the exact states defines a non-singular wave
operator $\Omega$ via $\ket{\Psi_i} = \Omega \ket{\psi^M_i}$. Assuming
that this wave operator has an inverse, the transformed Hamiltonian is
related to the full Hamiltonian by a similarity transformation
\begin{equation}\label{eq:ham}
  \mathcal{H} = \Omega^{-1}H\Omega\ .
\end{equation}

There is no unique representation for the wave operator $\Omega$.
Different choices for $\Omega$ may then give different results for the
perturbative expansion of the transformed interaction. Furthermore,
even the same defining equation for $\Omega$ can be solved by
different iterative schemes. As an example one often employs the
ansatz by Lee and Suzuki~\cite{lee80,suzuki80} for the wave operator

\begin{equation}
  \Omega = e^{\omega}
\end{equation}

where $\omega$ is known as the correlation operator. The correlation
operator generates the component of the wave function in the $Q$-space
and can be expanded in terms of Feynmann--Goldstone diagrams
\cite{Hjorth-Jensen1995}. One choice for the correlation operator is
\begin{equation}\label{eq:omega}
  \omega = Q\mathcal{\omega}P
\end{equation}
and thus we have $\Omega = 1+\omega$.

Applying the operator $Q$ on Eq.~(\ref{eq:psimodel}), one immediately
obtains $Q\mathcal{H}P=0$. Together with Eq.~(\ref{eq:ham}), this then
leads to the so-called decoupling equation
\begin{equation}\label{eq:als_dec}
  \omega P H Q \omega + \omega P H P - Q H Q \omega - QHP = 0
\end{equation}
which is to be solved in $\omega$ by iteration. For this non-linear
matrix equation no general solution method is known, nor the number of
existing solutions.

In this work we will discuss and generalize the solution of the
decoupling equation for arbitrary coupled channels. The construction
of the solution is based on the iteration method after Lee and
Suzuki~\cite{lee80,Lee1980} and Andreozzi~\cite{andreozzi}, labeled as
'ALS iteration' in the following.
\subsection{The ALS iteration}\label{subsec:ALS}

We start with an overview of the main ingredients of the standard ALS
iteration. As already mentioned, this is not the only existing method
which solves the decoupling equation.

Defining the model space effective Hamiltonian,
\begin{equation}
  p\left(\omega\right) = P \mathcal{H} P = PHP + PHQ\omega\ ,
\end{equation}
and its complement $Q$-space Hamiltonian,
\begin{equation}
  q\left(\omega\right) = Q \mathcal{H} Q = QHQ - \omega P H Q\ ,
\end{equation}
an iterative solution $\omega_{sol}\equiv\sigma$ of
Eq.~(\ref{eq:als_dec}) can be obtained by
\begin{equation}\label{eq:als_sum}
  \sigma = \sum_{n=0}^\infty x_n\ ,
\end{equation}
where the ${x_n}$'s satisfy successively
\begin{gather}
  x_0 = \frac{-1}{QHQ} QHP\ ,\nonumber\\
  x_1 = \frac{1}{q\left(x_0\right)} x_0 p\left(x_0\right)\ ,\nonumber\\
  \vdots \nonumber\\
  x_n = \frac{1}{q\left(x_0 +\cdots + x_{n-1}\right)} x_{n-1}
  p\left(x_0 + \cdots + x_{n-1}\right)\ .\nonumber
\end{gather}

Once a solution of these equations is known, the low-momentum
effective Hamiltonian can be expressed as
\begin{equation}\label{eq:hlowk}
  \hlowk = p(\sigma)\ .
\end{equation}
Subtracting from $\hlowk$ the kinetic energy in the model space, $PH_0
P$, this yields the wanted low-momentum potential $\vlowk$. As a
remark we note, that this iterative solution $\sigma$ is equivalent to
the LS vertex renormalization solution (see Refs.~\cite{Lee1980,lee80}).

Since $x_n \to 0$ for large $n$, only a finite number of terms in the
sum (\ref{eq:als_sum}) are needed numerically in order to reach a
desired precision. Furthermore, this condition implies for each
iteration step $n$ the relation
\begin{equation}\label{eq:als_conv}
  \left|\lambda_{p M}\right| < \left|\lambda_{q m}\right|\ ,
\end{equation}
where $\lambda_{p M}$ is the largest $P$-space eigenvalue (absolute
value) and $\lambda_{q m}$ the smallest $Q$-space eigenvalue. Thus, as
a necessary condition, this iteration method converges to the $d$
eigenvalues of the Hamiltonian of smallest absolute value.

A slight variation of the standard ALS iteration consists in the
subtraction of a constant multiple of the identity, $m_0 I$, from the
full Hamiltonian $H$. In this case the iteration will converge to the
shifted set of $d$ eigenvalues which are nearest to the chosen
constant $m_0$.
The choice of $m_0$ is not arbitrary. One has to make sure that the
eigenvalues closest to $m_0$ are really inside the $P$-space. For
numerical applications, this variation can also be used to accelerate
the iteration since the required shift of the eigenvalues is smaller
if $m_0$ is included.
Furthermore, calculating the $\vlowk$ for the hyperon-nucleon Nijmegen
potential ~\cite{rijken} (cf.~Ref.~\cite{1letter}) the constant shift
in the iteration is useful to avoid an unphysical bound state, which
is inherent in the $^1S_0$ isospin $3/2$ channel
(cf.~Ref.~\cite{miy1999}).

The important observation is so far that each of the iteration schemes
converges only to a set of $d$ consecutive eigenvalues of the
Hamiltonian, which are supposed to be ordered in a nondecreasing order.

Andreozzi described yet another generalization of the iteration
procedure that enables one to shift the eigenvalues individually by a
set of different numbers instead of just one constant multiple of the
identity (cf.~Ref.~\cite{Suzuki:1993pm}). If one denotes by $(A)_i$
the $i$-th column of a matrix $A$, the decoupling equation for this
case can be written in the form
\begin{eqnarray}\label{eq:mals}
  &\omega PHQ (\omega)_i + \omega\left(PHP-M\right)_i - \left(QHQ-m_i
    I_q\right)(\omega)_i& \nonumber \\
  &- \left(QHP\right)_i=0 \qquad \text{for}\  i=1,\ldots,d&
\end{eqnarray}
where the $d$-dimensional diagonal matrix $M = \diag
\left(m_1,\ldots,m_d\right)$ contains the arbitrarily chosen numbers
$m_i$. Thus, for each column there is a separate decoupling equation
to be solved. Using the preceding iteration procedure for the $i$-th
equation one obtains the solution
\begin{equation}
  \left(x_n\right)_i = \frac{1}{q\left(\sigma_{n-1}\right) -m_i I_q}
  x_{n-1}\left(p\left(\sigma_{n-1}\right)-M\right)_i\ .
\end{equation}

This modified ALS iteration scheme will converge to the eigenvalues
closest to the chosen set of numbers $m_i$.

This modification leads also to a stronger
convergence condition for each iteration step $n$
(cf.~Eq.~(\ref{eq:als_conv}))
\begin{equation}\label{eq:mals_conv}
  \left|\lambda_{p,i}-m_i\right| \leq \left|\lambda_{q,j} - m_i\right|
  \quad \forall\, i,j\ ,
\end{equation}
where $\lambda_{p,i}$ denotes the $i$-th eigenvalue in the model space
and $\lambda_{q,j}$ all eigenvalues in the $Q$-space. Thus, for a
fixed eigenvalue $i$ condition (\ref{eq:mals_conv}) there are $n-d$
inequalities for all eigenvalues in the $Q$-space.

All iteration methods, presented so far, converge rapidly for
channels with only one mass involved (one-flavor channel). For coupled
channels with different masses (multi-flavor channels) new phenomena
emerge and the convergence behavior (if any) of the iteration methods
changes dramatically as already indicated in Ref.~\cite{1letter}. This
becomes especially relevant e.g.~for the hyperon-nucleon interaction
in the particle basis, where up to four particles with different
masses are coupled.

%
%
\section{Multi-flavor Channels}

As already mentioned, the standard and modified ALS iteration methods
can fail for coupled channels interactions including different masses.
In this Section we will assess the reasons for this failure in detail.
Instead of considering the realistic hyperon-nucleon interaction with
different particles, we introduce a simple and solvable model which
mimics all qualitative and relevant features of the realistic
interaction. On the basis of this model the technical modifications of
the ALS iteration method needed as a remedy for the coupled
multi-flavor channels are discussed.

In the chosen model the failure of the standard and modified ALS
iteration methods for coupled multi-flavor channels and its cure are
seen immediately. We will restrict ourselves to a coupled two-flavor
channel in order to keep the discussion simple. The generalization to
arbitrary coupled multi-flavor channels with different masses is
straightforward.

\subsection{A simple model}

The model is governed by the Hamiltonian $H(k,k') =H_0(k) + V(k,k')$
which is in this case a $2\times2$ matrix in flavor-channel space. For
the kinetic energy of the $i$-th channel we use
\begin{equation}\label{eq:kinE}
  H_{0,i}(k)=\frac{k^2}{2\mu_i} + m_i\quad ; \quad i=1,2
\end{equation}
with $m_1 = 9$ and $m_2 =13$. All quantities in our model are
dimensionless. The reduced mass is defined by $\mu_i = m_0
m_i/(m_0+m_i)$ with a fixed mass $m_0 = 8$ for each $i$-th channel.
The mass parameters $m_i$ are adjusted in such a way that typical
convergence problems of the iteration method, emerging in a realistic
multi-flavor channel calculation, also appear in this model.

As a bare Hermitian interaction we choose a Fourier-transformed Yukawa
potential
\begin{equation}
V_{ij}(k,k') = \frac{a_{ij}}{b_{ij} k k'} \log \left(\frac{4 k k' +
    b_{ij}^2}{b_{ij}^2}\right)\ ,\  i,j=1,2 \ ;
\end{equation}
with the parameters $a_{11} = 4.3$, $a_{12} = 2.0$, $a_{22} = 3.7$ and
$b_{11} = -3.0$, $b_{12} = -8.0$, $b_{22} = -4.0$. The parameters
$a_{ij}$ reflect the depths and the $b_{ij}$ the effective ranges of
the corresponding potentials. They are chosen in such a way that the
diagonal potentials $V_{ii}$ are larger than the off-diagonal ones.

In the two-flavor channel, the block structure of the Hamiltonian looks
like
\begin{equation}
  H \equiv \begin{pmatrix}
    H_{11} & H_{12} \\
    H_{21} & H_{22}
  \end{pmatrix}= \begin{pmatrix}
    H_{0,1}+V_{11} & V_{12} \\
    V_{21} & H_{0,2}+V_{22}
  \end{pmatrix}\ ,
\end{equation}
with the relation $V_{21}(k,k') = V_{12}(k',k)$.

For each flavor-channel a $P$-space ($Q$-space) with two mesh points
and a cutoff $\Lambda_P =5$ ($\Lambda_Q=10$) is introduced. Here we
use only two mesh points in order to simplify our discussion of
convergence. In realistic calculations we use about 64 mesh points for
each channel.

Thus, each subblock $H_{ij}$ becomes now a $4\times 4$ matrix
\begin{equation}
  H_{ij} = \begin{pmatrix}
    P H_{ij} P & P H_{ij} Q \\
    Q H_{ij} P & Q H_{ij} Q
  \end{pmatrix}\quad ; \quad i,j=1,2\ .
\end{equation}

\begin{figure}
  \centering \includegraphics[width=0.95\linewidth]{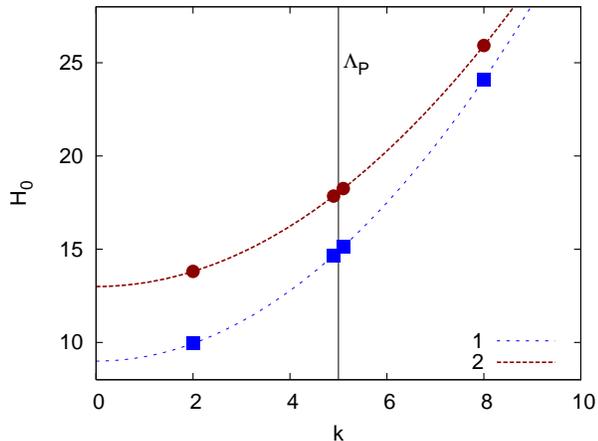}
  \caption{ The kinetic energies $H_{0,i}$ as function of the momentum $k$
    for the model. The line with squares denotes the first-flavor
    ($i=1$) channel, the other one the second-flavor ($i=2$) channel
    (see text for details). }
  \label{fig:toymodelcut1}
\end{figure}

In Fig.~\ref{fig:toymodelcut1} the kinetic energy Eq.~(\ref{eq:kinE})
as function of the momentum $k$ for the $i$-th flavor channel is
shown. Our choice of the $P$- and $Q$-space discretization yields four
mesh points on each curve (squares for the first channel and bullets
for the second channel). They are chosen in such a way that two mesh
points of each flavor channel are close to the $P$-space cutoff
$\Lambda_P$ which is depicted by a vertical line in the figure.

In the following we apply the standard and modified ALS iteration
methods, described in the previous Section, to this model. The
results are then compared with the exact solution of the
model. It is solved by numerical diagonalization, which produces the
following eight eigenvalues $H = \diag(9.55, 14.50, 15.05, 24.04,
13.64, 17.76, 18.20, 25.88)$. Since the Hamiltonian is dominated by
the kinetic energy contribution $H_0$, all eigenvalues are close to
these points shown in Fig.~\ref{fig:toymodelcut1}.

\subsection{ALS iteration methods}

We use two different quantities to monitor the convergence of the ALS
iteration methods. As a necessary convergence condition for all
iteration techniques the $x_n$ should tend to zero for $n\to \infty$.
We therefore use the norm $\norm{x_n}$ (maximal absolute column sum
norm of $x_n$) as function of the number of iterations $n$ as one
convergence criterion. Thus, if this norm increases, the iteration
method cannot converge.

At each iteration step $n$ the approximate solution
\begin{equation}
  \sigma_n = \sum_{i=0}^n x_i
\end{equation}
is plugged into the decoupling equation (cf.~Eq.~(\ref{eq:als_dec}))
\begin{equation}
  d_n = \sigma_n P H Q \sigma_n + \sigma_n P H P + Q H Q \sigma_n +
  QHP\ ,
\end{equation}
defining in this way the quantity $d_n$. Its norm $\norm{d_n}$
represents a deviation (distance) from the exact solution, where the
decoupling equation is exactly fulfilled. This quantity should also
tend to zero for $n\to\infty$.
As a second convergence criterion we calculate the eigenvalues at each
iteration step by diagonalization which is surely possible for the
model and monitor their behavior as function of the number of
iterations.

\begin{figure}
  \includegraphics[width=0.95\linewidth]{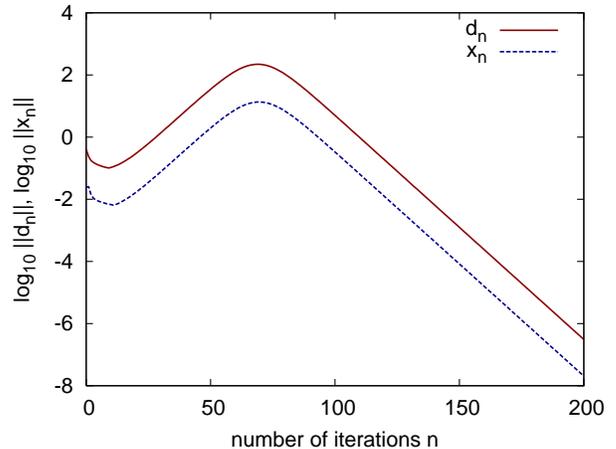}
  \caption{The norm of $x_n$ and of the deviation from zero $d_n$ of the
    decoupling equation for the model as function of the iteration
    number $n$ using the standard ALS iteration method with one
    momentum cutoff. }
  \label{fig:toymodelsALSit}
\end{figure}

We begin our analysis with the standard ALS iteration method as
described in Sec.~\ref{subsec:ALS}. In Fig.~\ref{fig:toymodelsALSit}
the norm $\norm{x_n}$ and the deviation $\norm{d_n}$ from the exact
solution of the decoupling equation are shown as a function of the
iteration number $n$. After the first few iterations, these quantities
increase with $n$ signaling a divergence of the method but then
decrease again for a larger number of iterations. This could lead to
the incorrect conclusion that the ALS iteration method converges anyway to
the right eigenvalues.

\begin{figure}
  \centering \includegraphics[width=0.95\linewidth]{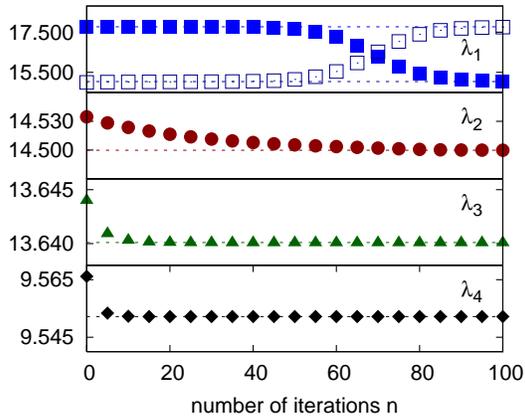}
  \caption{The $P$-space eigenvalues evolution for the model using the
    standard ALS iteration with one momentum cutoff. The open squares
    label the nearest $Q$-space eigenvalue. See text for details.}
  \label{fig:toymodelsALSe}
\end{figure}

This is demonstrated in the following: The eigenvalues in the
$P$-space are given by the eigenvalues of $p(\sigma_n)$ and are
plotted in Fig.~\ref{fig:toymodelsALSe} as a function of the iteration
number. The dashed lines are the final exact eigenvalues obtained by
diagonalization of the model Hamiltonian.

The starting values for the eigenvalues are
given by $P H P=p(0)$. Due to our grid choice for the model we can
identify the $P$- and $Q$-space eigenvalues uniquely. We violate in
this way the convergence condition (\ref{eq:als_conv}) because the
$Q$-space eigenvalue near the cutoff is smaller than the corresponding
$P$-space eigenvalue (cf.~Fig.~\ref{fig:toymodelcut1}). Thus, we
expect that the standard ALS iteration method does not converge.

This is clearly visible in Fig.~\ref{fig:toymodelsALSe} for the
eigenvalue $\lambda_1$ where the evolution of the $P$-space eigenvalue
(filled squared) and the corresponding $Q$-space eigenvalue (open
squared) are shown. Although the $x_n$'s tend to zero, the eigenvalues
converge for large $n$ to the ones with minimum absolute value. The
$P$-space eigenvalue evolves from $\lambda_1 \sim 17.78$ to $15.05$
and the $Q$-space one from $15.00$ to $17.76$. This mixing between a
$P$- and $Q$-space eigenvalue during the decimation is typical for a
divergence in the ALS iteration method. Due to our grid choice we have
only one eigenvalue mixing the other three $P$-space eigenvalues
converge to the correct solutions. But in general an arbitrarily larger
number of eigenvalues which violate condition (\ref{eq:als_conv})
could mix.

Unfortunately, for realistic interactions like the hyperon-nucleon
interaction it is not possible to identify and sort the proper $P$- and
$Q$-space eigenvalues for each $i$th-subblock. One has to find
alternative criteria which signal an eigenvalue mixing. One
alternative is the norms $\norm{x_n}$ and $\norm{d_n}$. During the
mixing the $\norm{x_n}$ and $\norm{d_n}$ rise and once the eigenvalue
sorting is completed they begin to drop again. Thus, a rising
$\norm{x_n}$ (or $\norm{d_n}$) and a following dropping with the
iteration number $n$ signals a level crossing and the standard ALS
iteration method converges to wrong results. This behavior is shown in
Fig.~\ref{fig:toymodelsALSit}.

As already mentioned, the standard ALS iteration method can be
modified and improved by adding a diagonal matrix $M = \diag
(m_1,\ldots)$ to the decoupling equation with arbitrary chosen $m_i$
(cf.~Eq.~(\ref{eq:mals}). Then, the eigenvalues converge closest to
the chosen numbers $m_i$. But the crucial point here is how to chose
this matrix $M$? No general criterion is known. For the model
calculation it is possible to try several different ans\"atze, which
indeed prevent the mixing of the $P$- and $Q$-space eigenvalues and
result in stable eigenvalues.

Unfortunately, all these ans\"atze cannot be applied to realistic
interactions. For a realistic coupled channel calculation much more
mesh points have to be taken into account. Then the identification of
the $P$- and $Q$-space eigenvalues for each flavor channel $i$ is not
possible any longer. This is also visible in
Fig.~\ref{fig:toymodelcut1}. One cannot distinguish between the
channel 2 $P$-space eigenvalues for small momenta $k$ and the channel
1 $P$-space eigenvalues for momenta around the cutoff $\Lambda_P$.
Even an alternative identification of the appropriate eigenvalues by
means of eigenvector distributions, as described e.g.~in
Ref.~\cite{Fujii04}, fails for realistic multi-flavor channel
interactions.

For our model the situation is different. By construction, the
eigenvalues of the $P$- and $Q$-space are well-separated and can be
assigned uniquely.
One way out of this dilemma is the introduction of a so-called
channel-dependent cutoff. The use of the channel-dependent cutoff will
cure not only the identification problem of the eigenvalues but also
has a natural physical interpretation which we will discuss in the
following.

\subsection{Channel-dependent cutoff}

\begin{figure}
  \centering \includegraphics[width=0.95\linewidth]{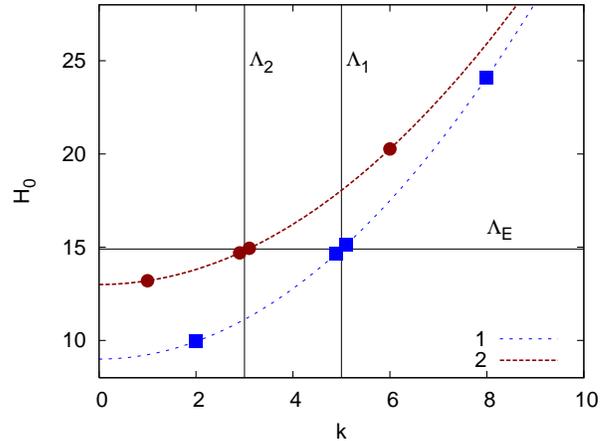}
  \caption{ The kinetic energy $H_0$ as a function of the momentum $k$ for
    the model similar to Fig.~\ref{fig:toymodelcut1}. The line with
    the squares denotes the first-flavor channel, the other one the
    second-flavor channel.
  }
  \label{fig:toymodelcut2}
\end{figure}

We assume that the dominant contribution to the full Hamiltonian comes
from its kinetic energy $H_0$ and the interaction is a small
perturbation.
The channel-dependent cutoff is a general prescription how to define
the model space for low-momentum interactions that include flavor
transitions with different masses. This is for example of relevance
for the hyperon-nucleon interaction in the particle basis~\cite{jletter}.
The idea of the channel-dependent cutoff is to truncate the kinetic
energy of each channel to the same energy $\Lambda_E$. The truncation
of the kinetic energy in this way leads to different momentum cutoffs
in each flavor channel $\Lambda_i$. This is depicted in
Fig.~\ref{fig:toymodelcut2}.

In principle there are two possibilities how to construct the momentum
cutoffs. One possibility is to fix the value of the kinetic energy
$\Lambda_E$ and then to calculate the corresponding momentum cutoffs
$\Lambda_i$. But the following procedure is better adapted to a
generalization of the $\vlowk$ method to multi-flavor channels.
Instead of fixing the kinetic energy $a$ $priori$, one calculates its
value for a given momentum cutoff of the flavor channel with the
smallest mass. For our model this means to choose $\Lambda_1$ for
channel 1 and calculate $\Lambda_2$ for channel 2 by requiring
\begin{equation}
  H_{0,1}\left(\Lambda_1\right) =
  H_{0,2}{\left(\Lambda_2\right)}\ .
\end{equation}
For example, choosing a momentum cutoff $\Lambda_1 = \Lambda = 5$ for
channel 1 this results in a smaller cutoff in channel 2
\begin{equation}
  \Lambda_2 = \sqrt{2 \mu_2 \left(\frac{\Lambda_1^2}{2 \mu_2}
      +m_1 - m_2\right)} \approx 3.1\,,
\end{equation}
which is denoted by the vertical line in Fig. \ref{fig:toymodelcut2}. Due
to the new momentum cutoff $\Lambda_2$ the mesh points have also
changed for channel 2. A generalization to multi-flavor channels is
straightforward.

Using this channel-dependent cutoff, the eigenvalues of the $P$- and
$Q$-space are now clearly separated and the necessary convergence
conditions (\ref{eq:als_conv}) and (\ref{eq:mals_conv}) are fulfilled.
Here, we have assumed that the potential is a perturbation and does
not influence the eigenvalue sorting strongly. We have checked this
assumption by a numerical calculation of the eigenvalues also in the
realistic calculation.

As already mentioned, note that with only one momentum cutoff for
multi-flavor channels it is not possible to fulfill these convergence
conditions and the ALS iteration methods must therefore fail.

Another, more physical motivation for the channel-dependent cutoff is
the following: The derivation of the $\vlowk$ renormalization group
equation is based on a Lippmann-Schwinger equation for the $T$-matrix.
The propagator in the LSE is inverse proportional to the energy, which
is truncated by a momentum cutoff.  For channels with different reduced
masses this would allow for different kinetic energies. Thus, it
is natural to restrict the kinetic energy available in the propagator
and not to restrict the momenta. For one-flavor channels this
difference does not play any role.

This becomes more transparent if one considers an on-shell scattering
process of particles with different masses. Assume the scattering of a
particle of channel $2$ into a particle of channel $1$ with a relative
incoming momentum around the model space cutoff, i.e.~$k' \approx
\Lambda$ and masses $m_2>m_1$. Since we consider on-shell scattering,
the energy
\begin{equation}
H_{0,2}(k'\approx\Lambda) \approx \frac{\Lambda^2}{2 \mu_2} + m_2=E\,.
\end{equation}
is fixed, say to $E$. This would result in
\begin{equation}
 E = H_{0,1}(k)=\frac{k^2}{2 \mu_1}+m_1 \ ,
\end{equation}
for the scattered particle $1$, which in turn yields for the outgoing
relative momentum
\begin{equation}
  k = \sqrt{\mu_1\left(\frac{\Lambda^2}{\mu_2} +2(m_2 -m_1) \right)}\ .
\end{equation}

Since we have assumed $m_2 > m_1$ this outgoing relative momentum is
larger than our cutoff $\Lambda$. For example, choosing for the
incoming momentum $k'=4.9$ the outgoing momentum would be $k\approx
7.4$. Since all momenta are limited by the cutoff $\Lambda =5$ such
transitions are not allowed. This is also clearly visible in
Fig.~\ref{fig:toymodelcut2}

Since the channel-dependent cutoff  restricts the kinetic energy it
can be represented by a horizontal line in this figure.
One never crosses this line in any on-shell scattering process.

When we apply the channel-dependent cutoff to the model, the mesh points
will change due to the different momentum cutoff, which is shown in
Fig.~\ref{fig:toymodelcut2}. This yields also a change of the eight
exact eigenvalues. In this case we find the eigenvalues $H =
\diag(9.55, 14.50, 15.05, 24.04, 12.96, 14.56, 14.88, 20.22)$ which
are again close to the points in Fig.~\ref{fig:toymodelcut2}.

\begin{figure}
  \includegraphics[width=0.95\linewidth]{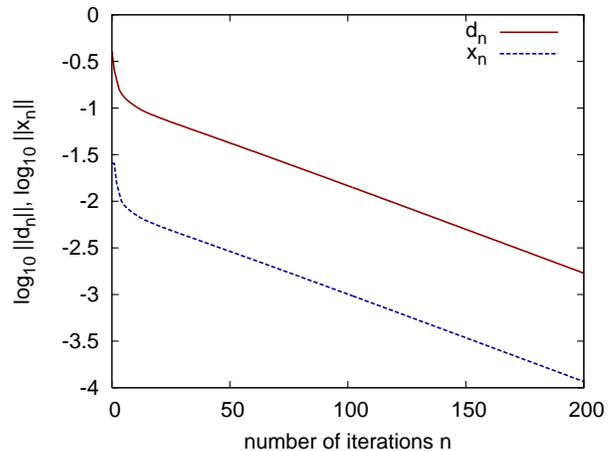}
  \caption{Same as Fig.~\ref{fig:toymodelsALSit} for the model using
    the standard ALS iteration method with a channel-dependent
    cutoff.}
  \label{fig:toymodel2sALSit}
\end{figure}

Due to the channel-dependent cutoff the convergence criterion for the
ALS iteration methods are fulfilled now because the $P$-space
eigenvalues are these eigenvalues which are closest to zero. The
different iteration schemes converge monotonic and to the correct
$P$-space eigenvalues. An example for the standard ALS iteration is
shown in Fig.~\ref{fig:toymodel2sALSit}.

\begin{figure}
  \centering \includegraphics[width=0.95\linewidth]{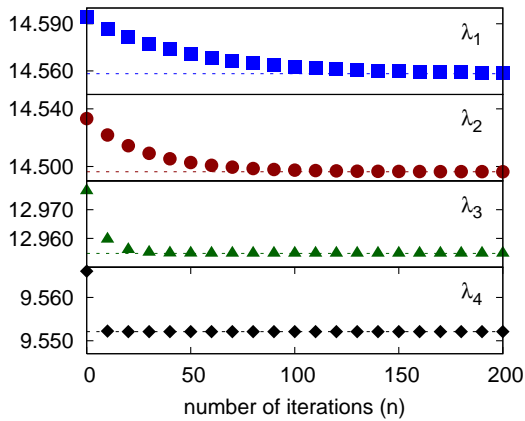}
  \caption{The $P$-space eigenvalue evolution for the model similar to
    Fig.~\ref{fig:toymodelsALSe} using the standard ALS iteration
    method with a channel-dependent cutoff.}
  \label{fig:toymodel2sALSe}
\end{figure}

The evolution of the four $P$-space eigenvalues for the model as a
function of the number of iterations is displayed in
Fig.~\ref{fig:toymodel2sALSe}. The dashed lines are the exact
eigenvalues. All eigenvalues converge to the correct ones. Due to the
small gap between the largest $P$-space and the smallest $Q$-space
eigenvalue in the model ,the number of iterations of is slightly larger
compared e.g.~to Fig.~\ref{fig:toymodelsALSe}.

%
%
\section{Summary}

In this work the ALS iteration methods and its modifications are
investigated, which are necessary to solve coupled-channel equations
with different masses. Instead of working with realistic interaction
potentials, a simple and solvable model, which mimics all the relevant
properties of a realistic interaction is introduced.

First, we applied the ALS iteration method and its modifications
to the one-flavor channel and discussed its failure for multi-flavor
channels. Due to the different masses inherent in the multi-flavor
case the necessary convergence conditions are violated.

By an introduction of a channel-dependent cutoff we could generalize
these ALS iteration modifications and could restore the convergence.
This channel-dependent cutoff results in an effective interaction which
is still energy-independent and thus preserves a major advantage of the
$\vlowk$ potential.

An on-shell scattering example of two particle with different masses
provides a more physical motivation for this novel cutoff. Without
the channel-dependent cutoff unphysical scattering processes out of the
model space become possible.
First encouraging results for a realistic hyperon-nucleon interaction
by means of the channel-dependent cutoff have been published in
Ref.~\cite{1letter}. Further calculations including a broader variety
of input potentials such as the Nijmegen potentials
(NSC97~\cite{rijken}, NSC89~\cite{PMMM89}), J{\"u}lich potential
(Juelich04~\cite{Haidenbauer:2005zh}), and potentials based on chiral
symmetry (LO potential ~\cite{Polinder:2006zh}) will be presented
elsewhere~\cite{jletter}.

\section*{Acknowledgments}

We thank A.~Schwenk and A.~Nogga for helpful discussions. This work
was supported in part by the U.S. DOE Grant No. DE-FG02-88ER/40388. MW
was supported by the BMBF grant 06DA116 and the FWF-funded Doctoral
Program ``Hadrons in vacuum, nuclei and stars'' at the Institute of
Physics of the University of Graz.


\end{document}